\documentclass[aps,prl,twocolumn,showpacs,floatfix]{revtex4}
\usepackage{epsfig}
\usepackage{bm}

\begin{document}
\title{Extraction of Plumes in Turbulent Thermal Convection}
\author{Emily S. C. Ching$^1$, H. Guo$^{1,2}$, 
Xiao-Dong Shang$^{1,3}$, P. Tong$^{3,4}$, and Ke-Qing Xia$^1$}
\affiliation{$^1$Department of Physics, The Chinese 
University of Hong Kong, Shatin, Hong Kong \\
$^2$State Key Laboratory for Studies of Turbulence and Complex Systems, 
Dept. of Mech.
and Engin. Sci., Peking University,
Beijing 100871, People's Republic of China \\
$^3$Department of Physics, Oklahoma State University, Stillwater,
Oklahoma 74078 \\
$^4$Department of Physics, Hong Kong University of Science
and Technology, Clear Water Bay, Kowloon, Hong Kong}

\date{\today}
 
\begin{abstract}
We present a scheme to extract information about plumes, a
prominent coherent structure in turbulent thermal convection,
from simultaneous local velocity and temperature measurements.
Using this scheme, we study the temperature
dependence of the plume velocity and understand the results
using the equations of motion. We further obtain the average 
local heat flux in the vertical direction at the cell center. 
Our result shows that heat is not mainly transported through the 
central region but instead through the regions near the sidewalls 
of the convection cell.
\end{abstract}
\maketitle

The Rayleigh-B{\'e}nard convection system 
consists of a closed cell of fluid heated
from below and cooled on the top.
The equations of motion, in Boussinesq approximation, are \cite{Landau}:
\begin{eqnarray}
&{\displaystyle \partial {\bf v} \over \displaystyle \partial t} 
+ {\bf v} \cdot {\bf \nabla} {\bf v}  
= -{\bf \nabla} p + \nu \nabla^2 {\bf v} + g \alpha \delta T {\bf \hat z}
\label{eqn1}\\
&{\displaystyle \partial T \over \displaystyle \partial t} 
+ {\bf v} \cdot {\bf \nabla} T = \kappa \nabla^2 T  
\label{eqn2} \\
&{\bf \nabla} \cdot {\bf v} = 0 
\label{eqn3}
\end{eqnarray}
where ${\bf v}$ is the velocity field, $p$ the pressure
divided by density, $T$ the temperature field, 
and ${\bf \hat z}$ is the unit vector in the vertical direction. 
Furthermore, $\delta T = T- T_0$ where 
$T_0$ is the mean temperature
of the bulk fluid,
$g$ is the acceleration due to gravity 
and $\alpha$, $\nu$, and $\kappa$ are respectively 
the volume expansion coefficient,
kinematic viscosity and thermal diffusivity of the fluid.
The state of fluid motion is characterized by the geometry of 
the cell and two dimensionless parameters: the Rayleigh 
number, Ra $= \alpha g \Delta L^3/ (\nu \kappa)$, which measures 
how much the fluid is driven and the Prandtl number, 
Pr $= \nu/\kappa$, which is the
ratio of the diffusivities of momentum and heat of the fluid.
Here $\Delta$ is the maintained temperature
difference between the bottom and the top,
and $L$ is the height of the cell.
When Ra is sufficiently large, the convective motion 
becomes turbulent.

In turbulent convection, local velocity and temperature 
measurements taken at a point within the
convection cell display complex fluctuations in time. On the
other hand, visualization of the flow reveals recurring 
coherent structures. One prominent coherent structure is a plume,
which is a mushroom-like flow generated by buoyancy.
Thus at least two strategies can be employed to study 
turbulent thermal convection or turbulent flows in general. One is to 
analyze and understand the fluctuations of 
the local measurements. The other is to characterize the
coherent structures and study and understand their dynamics. 
These two approaches are not independent but
provide complementary knowledge of turbulent flows. 
In particular, there is the natural question of whether
and how information about the coherent structures can be 
extracted from the local measurements. 

For turbulent flows not
driven by buoyancy, various methods including proper orthogonal 
decomposition, conditional sampling and wavelet analysis have been 
proposed to identify coherent vortical structures from local velocity 
measurements\cite{example}. On the other hand, much less 
work has been done in identifying plumes or extracting information
about plumes in turbulent thermal convection\cite{BL96,ZX02,JLMW99}. 
Belmonte and Libchaber\cite{BL96} used the skewness of the 
temperature derivative
as a signature of the plumes. Zhou and Xia\cite{ZX02} associated
the difference in the skewness of the positive and negative parts of the
temperature difference with the presence of plumes and identified the
plumes whenever the temperature difference becomes larger than a chosen
threshold\cite{MWAT01}. In Ref.\cite{JLMW99}, plumes are identified
when each of the temperature, vertical velocity or vertical vorticity 
is larger than some threshold.

In this Letter, we present a scheme to extract
information about plumes using simultaneous local velocity and 
temperature measurements. Our method makes explicit use of the
physical intuition that plumes are generated by buoyancy and thus
the velocity of a plume should correlate with the 
temperature fluctuation. Using this method,
we obtain the temperature dependence of the 
plume velocity and 
the average local heat flux in the
vertical direction at the cell center. 

The experimental measurements that we use
to illustrate and test our scheme of plume extraction were
taken in an aspect-ratio-one
cylindrical cell of  height
$L=20.5$~cm and filled with water\cite{SQTX03}. 
The velocity ${\bf v}(t)$ was measured
using a two-component laser Doppler velocimetry (LDV) system\cite{QT0102} 
while the temperature $T(t)$ was measured using a small movable
thermistor of 0.2~mm in diameter, 15~ms in time constant, and 
1~mK/$\Omega$ in temperature sensitivity. The velocity and temperature 
measurements were simultaneously taken
using a multichannel LDV interface module to synchronize the 
data acquisition \cite{SQTX03}.
A triggering pulse from the LDV signal processor
initiates the acquistion of an analog temperature signal.
The spatial separation between the LDV focusing spot and the thermistor
tip is kept at a minimal value of $0.7 \pm 0.2$~mm, a distance 3 times
larger than the tip diameter of the thermistor but 20 times smaller than
the correlation length between the temperature and velocity
fluctuations\cite{tobe}. The simultaneous velocity and temperature 
measurements can thus be assumed to be taken at the same location.
In this work, we study 
measurements taken at two locations within the convection cell: the 
cell center and near the sidewall (on the mid-plane, at 8~mm from the 
sidewall).  Near the sidewall,
the vertical velocity component $v_z$ and one horizontal velocity
component $v_y$, which is out of the rotation plane of the mean
large-scale circulation, were measured. 
At the cell center, all three velocity
components $v_x$, $v_y$ and $v_z$ were measured with
$v_x$ along the direction of the large-scale mean circulation 
near the bottom plate.
The number of velocity or temperature 
measurements at each location is $2-4 \times 10^5$, 
corresponding to 8 hours of real-time measurements.

From the simultaneous velocity and temperature measurements, we 
calculate $\langle {\bf v} | T(t)\rangle$, 
the conditional average of the velocity on the temperature 
measured at time $t$. In calculating 
$\langle {\bf v} | T(t)\rangle$, we take the average of
those velocity measurements with the
corresponding temperature measurements 
falling within $T(t) \pm 0.005^\circ$C.
Then we decompose the velocity measurement ${\bf v}(t)$ into 
a sum of $\langle {\bf v} | T(t) \rangle$ and the 
remaining part, denoted as ${\bf v_b}(t)$:
\begin{equation}
{\bf v}(t) = \langle {\bf v} | T(t) \rangle + {\bf v_b} (t)
\end{equation}
By construction, ${\bf v_b}(t)$ is uncorrelated with $T(t)$ and
averages to zero over time. Thus we take it as
the background velocity fluctuation. 
If ${\bf v}(t)$ correlates with $T(t)$, 
$\langle {\bf v} | T(t) \rangle$
is different from the ordinary average velocity 
$\langle v \rangle$ and is a function of $T$ as well as $t$
since $T(t)$ depends on $t$. In this case,
we define the velocity of the plume by ${\bf v_p}(t) \equiv 
\langle {\bf v}| T(t) \rangle$. In other words, ${\bf v_p}(t)$ 
is the part of ${\bf v}(t)$ that correlates with $T(t)$. 
If ${\bf v}(t)$ is uncorrelated with $T(t)$, 
${\bf v_b}(t) =  
{\bf v}(t) - \langle {\bf v} \rangle$
and no plume is present.  With this velocity 
decomposition, we can easily identify the local heat flux 
carried by the plumes.
The normalized local heat flux is given by 
${\bf j}(t) = [{\bf v}(t) \delta T(t)]L/(\kappa \Delta)$.
The part carried by the plumes is thus
${\bf j_p}(t) \equiv [{\bf v_p}(t) \delta T(t)] L/(\kappa \Delta)$.
When the contribution by the plumes is subtracted, the remaining
is the heat flux carried by the background velocity fluctuation,
which is ${\bf j_b}(t) = [{\bf v_b}(t) \delta T(t)] L/(\kappa \Delta)$ and
averages to zero over time.

It is found that the horizontal velocity components have 
little correlation with the temperature fluctuation, having
a correlation coefficient of ~0.1 and ~0.2 respectively 
at the cell center and near the sidewall. 
On the other hand, the vertical velocity component $v_z$
shows stronger correlation with $T(t)$, 
having a correlation coefficient of ~0.3 and ~0.5 respectively
at the cell center and near the sidewall. 
Consistently, we find that the difference between $\langle {\bf v} | T(t)
\rangle$ and $\langle {\bf v } \rangle $ lies mainly in the 
vertical direction. That is, 
${\bf v_p}(t) \approx v_{pz}(t) {\bf \hat{z}}$.

\begin{figure}
\centering
\includegraphics[width=.45\textwidth]{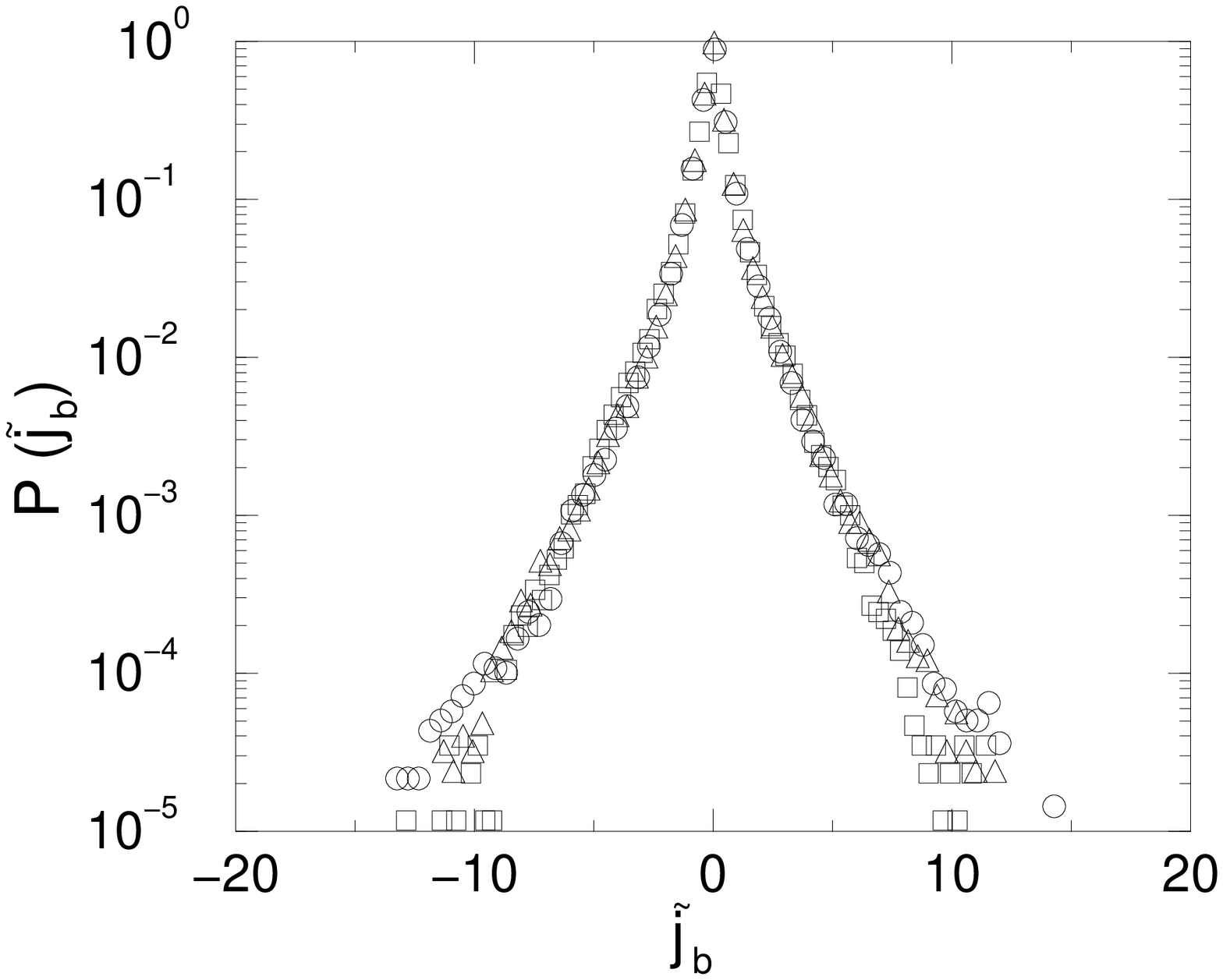}
\includegraphics[width=.45\textwidth]{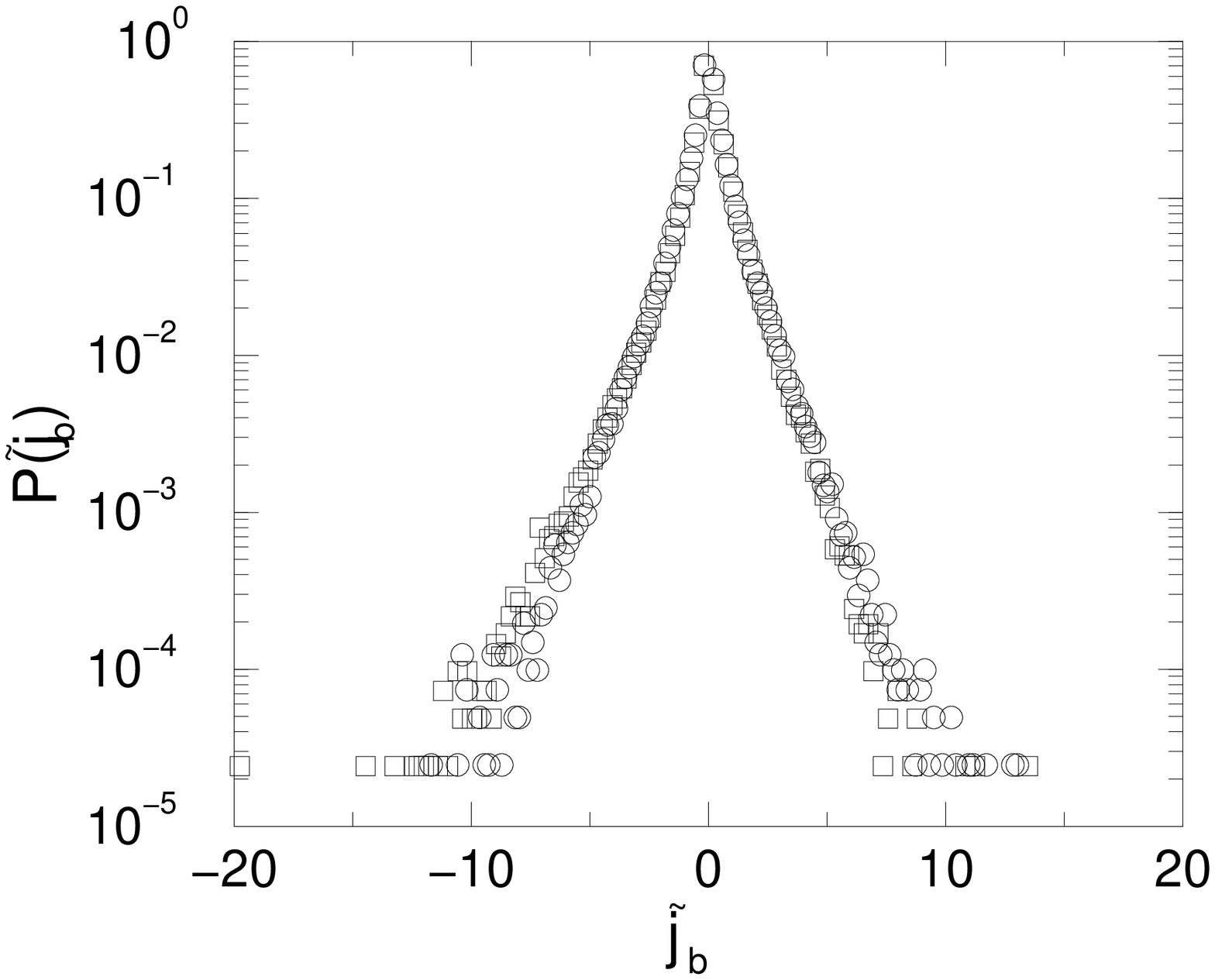}
\caption{The standardized probability density functions 
$P(\tilde j_{bx})$~(triangles),
$P(\tilde j_{by})$~(squares), and $P(\tilde j_{bz})$~(circles)
at Ra = $2.6 \times 10^9$. The measurements were made at the 
cell center (top panel) and 
near the sidewall (bottom panel).}
\label{fig2}
\end{figure}

It was recently found \cite{SQTX03} that the 
probability distribution of the local heat flux in the 
horizontal direction is approximately symmetric whereas 
that in the vertical direction is
skewed towards the positive fluctuation. 
This asymmetry in the vertical flux 
distribution was interpreted to be 
caused by the plumes\cite{SQTX03}. 
We use these results to test our scheme of 
plume extraction. We expect that after subtracting the 
the contribution by the plumes, the probability distributions of the
local heat flux in the horizontal and vertical
directions would become identical. 
In Fig.~\ref{fig2}, we show the probability distributions of 
$\tilde j_{bx}$, $\tilde j_{by}$, and $\tilde j_{bz}$.
Here the standardized variable 
$\tilde X$ is defined as $(X- \langle X \rangle)/\sigma_X$
where $\langle X \rangle $ and $\sigma_X$ are respectively
the mean and standard deviation of $X$.
In the calculation, we estimate the mean temperature of the bulk
fluid $T_0$ by the average temperature
measured at the cell center. 
As expected, the standardized distributions of the 
horizontal and vertical components of $\tilde {\bf j_b}$ are the same. 
Moreover, they are approximately the same at the two locations studied.

\begin{figure}
\centering
\includegraphics[width=.45\textwidth]{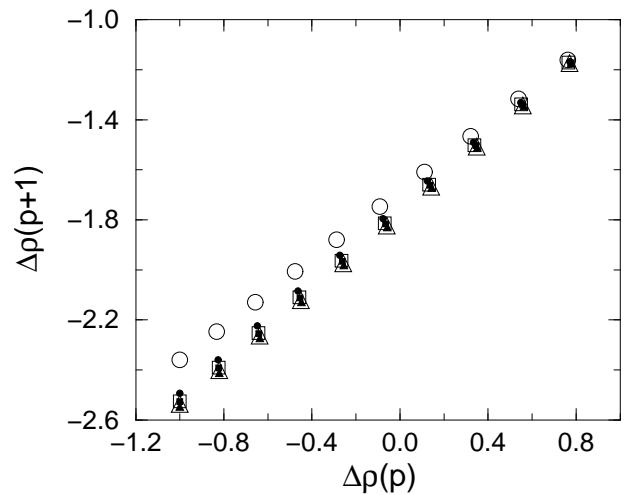}
\caption{$\Delta \rho (p+1)$ vs $\Delta \rho (p)$ for the
$x$-component~(triangles),
$y$-component~(squares), and
$z$-component~(circles) for ${\bf v}$~(open symbols) and
${\bf v_b}$~(filled symbols) at the cell center for Ra
$=4.8 \times 10^9$.}
\label{fig3}
\end{figure}

In a recent study\cite{CLQT03}, 
the three velocity components at the cell center 
were found to possess the She-Leveque hierarchical structure\cite{SL94}:
\begin{equation}
{S_{p+2}(\tau) \over S_{p+1}(\tau)} = A_{p} 
\left[{S_{p+1}(\tau) \over S_{p}(\tau)}\right]^{\beta}
[S^{(\infty)}(\tau)]^{1-\beta}
\label{SLHS}
\end{equation}
where $S_p(\tau) \equiv 
\langle |v(t+\tau)-v(t)|^p \rangle$ with $v$
being one of the three velocity components,
$A_{p}$ are constants independent
of $\tau$ and $S^{(\infty)}(\tau) \equiv \lim_{p \to \infty} 
{S_{p+1}(\tau)/S_{p}(\tau)}$.
Here, $\beta$ is a parameter whose value lies between $0$ and $1$.
The smaller $\beta$ is, the
more intermittent the velocity fluctuation is.
It was found that
the two horizontal velocity components are characterized
by the same value of $\beta$ but the
vertical velocity component is characterized by a 
smaller value of $\beta$. This distinction was attributed\cite{CLQT03}
to the presence of the plumes which makes the vertical
velocity component more intermittent.  
As a further test of our scheme of
plume extraction, we analyze 
${\bf v_b}(t)$ at the cell center to check whether its horizontal 
and vertical components will now be characterized by the same value
of $\beta$. To get the value of $\beta$, we follow the procedure
developed in Ref.\cite{ChingESS} and calculate the relative
scaling exponents $\rho (p)$ 
of the normalized structure functions, defined by
\begin{equation}
{S_p(\tau) \over S_2(\tau)^{p/2}} \sim
\left[{S_1(\tau) \over S_2(\tau)^{1/2}}\right]^{\rho(p)}
\label{rho}
\end{equation}
If Eq. (\ref{SLHS}) holds, we have \cite{ChingESS}
$\Delta \rho(p+1) = \beta \Delta \rho(p) - (1 + \beta)$,
where $\Delta \rho (p) \equiv \rho (p+1)-\rho(p)$.
In Fig.~\ref{fig3}, we plot $\Delta \rho(p+1)$ vs $\Delta \rho(p)$.
It can be seen that the data points for the
three velocity components of ${\bf v_b}(t)$ nearly fall
on the same line, showing that the
difference in the $\beta$-values of the horizontal and vertical
components indeed becomes vanishingly small.

\begin{figure}
\centering
\includegraphics[width=.45\textwidth]{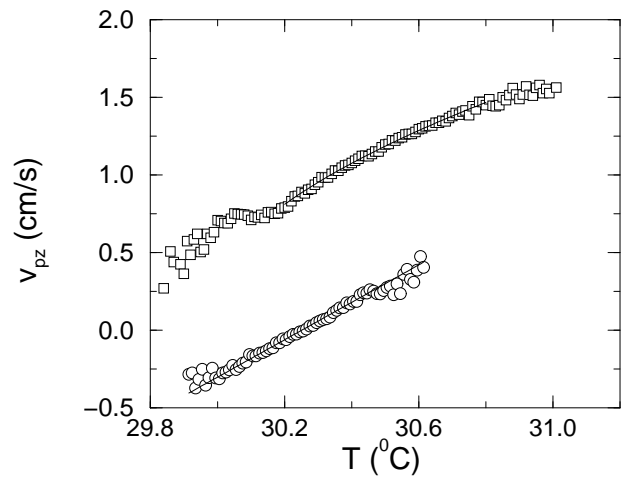}
\caption{$v_{pz}$ vs $T$ at the cell center~(circles) and
near the sidewall~(squares) at Ra $=2.6 \times 10^9$. The solid lines
from bottom to top are respectively the fits of
$a\delta T$ and $b\sqrt{T-T_s}$ for some constants $a$ and $b$.
The fitted value for $T_s$ is
$29.9^\circ C$ while $T_0$ is $30.2^\circ C$.}
\label{fig4}
\end{figure}

Next we use our scheme of plume extraction
to study the temperature dependence of the plume velocity.
The results are shown in Fig.~\ref{fig4}. We 
find that $v_{pz} = a \delta T$ 
at the cell center and $v_{pz} = b\sqrt{T-T_s}$ 
for $T > T_0 (> T_s)$ near the
sidewall for some constants $a$ and $b$. 
The fitted value of $T_s$ differs from the value of $T_0$ as estimated
by the average temperature at the cell center by about 1\%. 
We believe that this 
difference is caused by 
drifts in the mean temperature of the bulk fluid
over time and that the measurements at the two locations
were not taken at the same time. Hence, it is reasonable to 
conclude that $v_{pz} \propto \sqrt{\delta T}$ for 
$\delta T > 0$ near the sidewall.

\begin{figure}
\centering
\includegraphics[width=.45\textwidth]{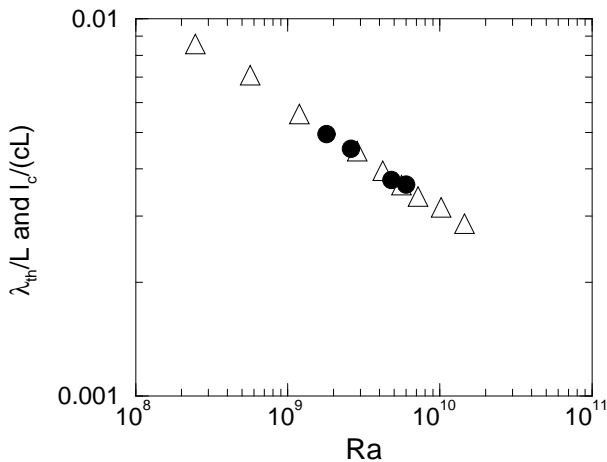}
\caption{$\lambda_{th}/L$ (triangles) \cite{LuiXia}
and $l_c/(cL)$ (filled circles)
vs Ra with $c=1.9$.}
\label{fig5}
\end{figure}

In the following, we shall understand this temperature dependence. 
At the cell center, the mean velocity vanishes and thus we 
expect the viscous term to dominate over the advection term. By 
balancing the
viscous term by the buoyancy term in Eq.~(\ref{eqn1}), we have
\begin{equation}
{\rm center:} \qquad {\nu v_{pz} \over l_c^2} = g \alpha \delta T 
\label{vpcenter}
\end{equation}
for some length scale $l_c$, which explains the temperature dependence
observed at the cell center.
Since plumes are excitations from the thermal boundary
layer, we expect $l_c$ to be of the order of the thermal boundary layer thickness
$\lambda_{th}$. We calculate $l_c$ from the fitted value of $a$ and 
compare $l_c/L$  with $\lambda_{th}/L$\cite{LuiXia}. As shown in Fig.~\ref{fig5},
$l_c/L$, after scaled down by a factor of about 2, coincides
with $\lambda_{th}/L$ over the limited Ra range studied.

Near the sidewall, the average
velocity of the plumes is equal to the average vertical velocity,
which is essentially the mean large-scale circulation velocity and
is relatively large.
Thus the advection term should dominate over the viscous term.
For $\delta T > 0$, we equate the advection term by the buoyancy term 
in Eq.~(\ref{eqn1}) and get
\begin{equation}
{\rm sidewall:} \qquad 
{v_{pz}^2 \over l_s } = g \alpha \delta T 
\qquad {\rm for} \ \delta T > 0 
\label{vpside}
\end{equation}
for some other length scale $l_s$. Hence $v_{pz} \propto \sqrt{\delta T}$ for
$\delta T > 0$ reproducing the temperature dependence observed near the
sidewall. Since the advection term is produced by 
the mean large-scale circulation, it seems 
reasonable to expect $l_s$ to be of the order of
the cell height $L$. Indeed $l_s$, calculated from the fitted value of $b$, 
is about $0.3-0.4 L$ and has little dependence on Ra. 

We further obtain the average local heat flux in the 
vertical direction at the cell center, $\langle j_z(t) \rangle_c$,
which can be thought of as a pointwise Nusselt number at the cell center. 
Since $\langle {\bf j_b}(t)\rangle = 0$, 
$\langle j_z(t) \rangle =
\langle j_{pz}(t) \rangle = 
\langle v_{pz}(t) \delta T(t) \rangle
L/(\kappa \Delta)$. Using 
(\ref{vpcenter}), we have
\begin{equation}
\langle j_{z}(t) \rangle_c = {\rm Ra}
 {\langle (\delta T)^2 \rangle_c \over \Delta^2} 
\left({l_c \over L}\right)^2
\label{jzcenter}
\end{equation}
Note that $\Delta_c \equiv \sqrt{\langle (\delta T)^2 \rangle_c}$ is 
the root-mean-square temperature fluctuation at the cell center. 
It was found that $\Delta_c/\Delta
\sim$  Ra$^{-1/7}$ both in low temperature 
helium gas\cite{Castaing} and in water\cite{XiaLui}. 
As $l_c/L$ goes like $\lambda_{th}/L$ 
(see Fig. \ref{fig5}) and $\lambda_{th}/L \sim $
Ra$^{-2/7}$\cite{LuiXia}, we have 
$\langle j_z(t) \rangle_c \sim$ Ra$^{1/7}$.
The scaling exponent 1/7 is clearly different from the
scaling exponent of approximately 2/7 of the Nusselt number\cite{GL}. 
This result thus shows explicitly that 
heat cannot be mainly transported
through the central region of the convection cell. 

In summary, we have presented a scheme to extract information about
plumes, the prominent coherent structures in turbulent thermal
convection. Our scheme involves a decomposition of the local velocity
into two parts. The part that correlates with the local
temperature measured at the same time is defined as
the plume velocity. Using this scheme of plume extraction, 
we have found the
temperature dependence of the plume velocity at the center 
and near the sidewall of the convection cell,
and understood such dependence from the
equations of motion. Moreover, we have obtained 
the average local heat flux in the vertical direction at
the cell center, and found that it has a scaling dependence 
with Ra different from that of 
the Nusselt number. This difference shows
that heat is not mainly transported through 
the central region but instead through
regions near the sidewalls of the convection cell. 
Further results about the mean large-scale circulation velocity 
and the velocity fluctuation at the cell center can be derived 
from Eqs. (\ref{vpcenter}) and (\ref{vpside}), and 
will be reported elsewhere.

The work at the Chinese University of Hong Kong 
was supported by the Hong Kong Research Grants Council (ESCC and HG by 
CUHK 4046/02P and XDS and KQX by CUHK 4242/01P).
PT was supported in part by NSF-0071823 and HKUST 603003.


\begin{thebibliography}{99}
\bibitem{tobe} The correlation length was obtained by varying the
separation between the temperature and velocity probes.
\end{thebibliography}
\end{document}